\RequirePackage{ifpdf}
\ifpdf 
\documentclass[pdftex]{sigma}
\else
\documentclass{sigma}
\fi

\begin{document}
\renewcommand{\PaperNumber}{082}

\FirstPageHeading

\renewcommand{\thefootnote}{$\star$}

\ShortArticleName{The Form Factor Program}

\ArticleName{The Form Factor Program: \\
a Review and New Results -- \\
the Nested $\boldsymbol{SU(N)}$ Of\/f-Shell Bethe
Ansatz\footnote{This paper is a contribution to the Proceedings of
the O'Raifeartaigh Symposium on Non-Perturbative and Symmetry
Methods in Field Theory
 (June 22--24, 2006, Budapest, Hungary).
The full collection is available at
\href{http://www.emis.de/journals/SIGMA/LOR2006.html}{http://www.emis.de/journals/SIGMA/LOR2006.html}}}

\Author{Hratchya M.  BABUJIAN~$^\dag$, Angela FOERSTER~$^\ddag$
and Michael KAROWSKI~$^\S$}

\AuthorNameForHeading{H.M. Babujian, A. Foerster and M. Karowski}

\Address{$^\dag$~Yerevan Physics Institute, Alikhanian Brothers 2,
Yerevan, 375036, Armenia}
\EmailD{\href{mailto:babujian@physik.fu-berlin.de}{babujian@physik.fu-berlin.de}}

\Address{$^\ddag$~Instituto de F\'{\i}sica da UFRGS, Av. Bento
Gon\c{c}alves 9500, Porto Alegre, RS - Brazil}
\EmailD{\href{mailto:angela@if.ufrgs.br}{angela@if.ufrgs.br}}

\Address{$^\S$~Theoretische Physik, Freie Universit\"{a}t Berlin,
Arnimallee 14, 14195 Berlin, Germany}
\EmailD{\href{mailto:karowski@physik.fu-berlin.de}{karowski@physik.fu-berlin.de}}
\URLaddressD{\href{http://www.physik.fu-berlin.de/~karowski/}{http://www.physik.fu-berlin.de/\~{}karowski/}}

\ArticleDates{Received September 29, 2006, in f\/inal form
November 16, 2006; Published online November 23, 2006}

\Abstract{The purpose of the ``bootstrap program'' for integrable
quantum f\/ield theories in $1+1$ dimensions is to construct
explicitly a model in terms of its Wightman functions. In this
article, this program is mainly illustrated in terms of the
sinh-Gordon model and the $SU(N)$ Gross--Neveu model. The nested
of\/f-shell Bethe ansatz for an $SU(N)$ factorizing S-matrix is
constructed. We review some previous results on sinh-Gordon form
factors and the quantum operator f\/ield equation. The problem of
how to sum over intermediate states is considered in the short
distance limit of the two point Wightman function for the
sinh-Gordon model.}


\Classification{81T08; 81T10; 81T40}

\section{Introduction}

The \textit{bootstrap program} to formulate particle physics in
terms of the scattering data, i.e.\ in terms of the S-matrix goes
back to Heisenberg \cite{Heis} and Chew \cite{Chew}. Remarkably,
this approach works very well for integrable quantum f\/ield
theories in $1+1$ dimensions \cite{STW,KTTW,KT,ZZ,KW,BKW}. The
program does \emph{not} start with any classical Lagrangian.
Rather it classif\/ies integrable quantum f\/ield theoretic models
and in addition provides their explicit exact solutions in term of
all Wightman functions. We achieve contact with the classical
models only, when at the end we compare our exact results with
Feynman graph (or other) expansions which are usually based on
Lagrangians. However, there is no reason that the resulting
quantum f\/ield theory is related to a classical Lagrangian.

\renewcommand{\thefootnote}{\arabic{footnote}}
\setcounter{footnote}{0}

One of the authors (M.K.) et al.~\cite{KTTW} formulated the
on-shell program i.e.\ the exact determination of the scattering
matrix using the Yang--Baxter equations. The concept of
generalized form factors was introduced by one of the authors
(M.K.) et al.~\cite{KW}. In this article consistency equations
were formulated which are expected to be satisf\/ied by these
quantities. Thereafter this approach was developed further and
studied in the context of several explicit models by
Smirnov~\cite{Sm} who proposed the form factor equations
$(i)$--$(v)$ (see below) as extensions of similar formulae in the
original article \cite{KW}. These formulae were then proven by two
of the authors et al.~\cite{BFKZ}. In the present article we apply
the form factor program for an $SU(N)$ invariant S-matrix
(see~\cite{BFK1}). The procedure is similar to that for the $Z(N)$
and $A(N-1)$ cases \cite{BK04,BFK} because the bound state fusions
are similar in these three models. However, the algebraic
structure of the form factors for the $SU(N)$ model is more
intricate, because the S-matrix also describes backward
scattering. We have to apply the nested
``of\/f-shell''\footnote{``Of\/f-shell'' in the context of the
Bethe ansatz means that the spectral parameters in the algebraic
Bethe ansatz state are not f\/ixed by Bethe ansatz equations in
order to get an eigenstate of a Hamiltonian, but they are
integrated over.} Bethe ansatz, which was originally formulated by
one of the authors (H.B.) \cite{B1,B3,BF} to calculate correlation
function in WZNW models (see also \cite{FR,SV}).

Finally the Wightman functions are obtained by taking integrals
and sums over intermediate states. The explicit evaluation of all
these integrals and sums remains an open challenge for almost all
models, except the Ising model \cite{Ba,MTW,SMJ,BKW,YZ}. In this
article we discuss this problem for the examples of the
sinh-Gordon model (see \cite{BK5}). We investigate the short
distance behavior of the two-point Wightman function of the
exponentiated f\/ield.

\section{The ``bootstrap program''}

The `bootstrap program' for integrable quantum f\/ield theories in
$1+1$-dimensions provides the solution of a model in term of all
its Wightman functions. The result is obtained in three steps:

\begin{enumerate}\itemsep=0pt
\item The S-matrix is calculated by means of general properties
such as unitarity and crossing, the Yang--Baxter equations (which
are a consequence of integrability) and the additional assumption
of `maximal analyticity'. This means that the two-particle
S-matrix is an analytic function in the physical plane (of the
Mandelstam variable $(p_{1}+p_{2})^{2}$) and possesses only those
poles there which are of physical origin. The only input which
depends on the model is the assumption of a particle spectrum with
an underlining symmetry. Typically there is a correspondence of
fundamental representations with multiplets of particles.
A~\emph{classification} of all S-matrices obeying the given
properties is obtained.

\item Generalized form factors which are matrix elements of local
operators
\[
^{\rm out}\left\langle
\,p_{m}^{\prime},\ldots,p_{1}^{\prime}\left\vert
\mathcal{O}(x)\right\vert p_{1},\ldots,p_{n}\,\right\rangle ^{\rm
in}\,
\]
are calculated by means of the S-matrix. More precisely, the
equations $(i)$--$(v)$ as listed in Section~\ref{s3} are solved.
These equations follow from LSZ-assumptions and again the
additional assumption of `maximal analyticity'~\cite{BFKZ}.

\item The Wightman functions are obtained by inserting a complete
set of intermediate states. In particular the two point function
for a Hermitian operator $\mathcal{O}(x)$ reads
\begin{gather*}
\langle\,0\left\vert \mathcal{O}(x)\,\mathcal{O}(0)\right\vert
0\,\rangle
=\sum_{n=0}^{\infty}\frac{1}{n!}\idotsint\frac{dp_{1}\cdots dp_{n}}%
{\,(2\pi)^{n}2\omega_{1}\cdots 2\omega_{n}}\\
\phantom{\langle\,0\left\vert
\mathcal{O}(x)\,\mathcal{O}(0)\right\vert 0\,\rangle=}{}
\times\left\vert \left\langle \,0\left\vert
\mathcal{O}(0)\right\vert p_{1},\ldots,p_{n}\,\right\rangle
^{\rm in}\right\vert ^{2}e^{-ix\sum p_{i}}.
\end{gather*}
Up to now a direct proof that these sums converge exists only for
the scaling Ising model \cite{Ba,MTW,SMJ} and the non-unitary
`Yang--Lee' model \cite{Sm3}.
\end{enumerate}

Recently, Lechner \cite{Le} has shown that models with factorizing
S-matrices exist within the framework of algebraic quantum
f\/ield theory. For the algebraic approach to quantum f\/ield
theories with factorizing S-matrices see the works of Schroer and
Schroer--Wiesbrock \cite{Sch1,Sch2,SW}. A determinant
representation for correlation functions of integrable models has
been obtained by Korepin et al.~\cite{KO,KoSl}.

\subsection*{Integrability}

Integrability in (quantum) f\/ield theories means that there exist
inf\/initely many local conservation laws
\[
\partial_{\mu}J_{L}^{\mu}(t,x)=0\qquad(L=\pm1,\pm3,\dots).
\]
A consequence of such conservation laws in 1+1 dimensions is that
there is no particle production and the n-particle S-matrix is a
product of 2-particle S-matrices
\[
S^{(n)}(p_{1},\dots,p_{n})=\prod_{i<j}S_{ij}(p_{i},p_{j}).
\]
If backward scattering occurs the 2-particle S-matrices will not
commute and one has to specify the order. In particular for the
3-particle S-matrix there are two possibilities
\begin{gather*}
\mathbf{S^{(3)}=S_{12}S_{13}S_{23}=S_{23}S_{13}S_{12}}\\[2mm]%
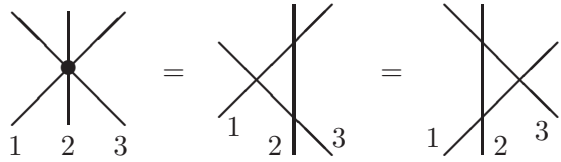
\begin{array}
[c]{c}%
\unitlength5mm\begin{picture}(17,4)(-6,0) \thicklines
\put(-5.5,.8){\line (1,1){3}} \put(-4,.8){\line(0,1){3}}
\put(-2.5,.8){\line(-1,1){3}}
\put(-4,2.3){\makebox(0,0){\Large$\bullet$}} \put(-5.6,0){$1$}
\put (-4.2,0){$2$} \put(-2.8,0){$3$} \put(-1.5,2){$=$}
\put(0,1){\line(1,1){3}} \put(0,3){\line(1,-1){3}}
\put(2,0){\line(0,1){4}} \put(4.3,2){$=$} \put(6,0){\line(1,1){3}}
\put(6,4){\line(1,-1){3}} \put(7,0){\line(0,1){4}}
\put(.2,.5){$1$} \put(1.3,0){$2$} \put(3,.2){$3$}
\put(5.5,.2){$1$} \put(7.3,0){$2$} \put(8.4,.4){$3$} \end{picture}
\end{array}
\end{gather*}
which yield the \textbf{``Yang--Baxter Equation''.}

\paragraph{The two particle S-matrix}\ is of the form%
\[
S_{\alpha\;\beta}^{\beta^{\prime}\alpha^{\prime}}(\theta_{12})=%
\begin{array}
[c]{c}%
\unitlength2mm\begin{picture}(5,6.3) \thicklines
\put(0,1){\line(1,1){4}} \put(2,3){\makebox(0,0){$\bullet$}}
\put(4,1){\line(-1,1){4}} \put (0,-.5){$\alpha$}
\put(3.5,-.5){$\beta$} \put(3.5,5.8){$\alpha'$}
\put(0,5.8){$\beta'$} \end{picture}
\end{array}
\]
where $\alpha$, $\beta$ etc denote the type of the particles and
the rapidity
dif\/ference $\theta_{12}=\theta_{1}-\theta_{2}>0$ is def\/ined by $p_{i}%
=m_{i}(\cosh\theta_{i},\sinh\theta_{i})$. We also use the short
hand notation $S_{12}(\theta_{12})$. It satisf\/ies unitarity
\begin{equation}
S_{21}(\theta_{21})S_{12}(\theta_{12})=1:~~~~~%
\begin{array}
[c]{c}%
\unitlength3mm\begin{picture}(8,5) \put(0,1){\line(1,1){2}}
\put(2,1){\line(-1,1){2}} \put(0,3){\line(1,1){2}}
\put(2,3){\line(-1,1){2}} \put(6,1){\line(0,1){4}}
\put(8,1){\line(0,1){4}} \put(3.7,2.7){$=$} \put(0,-.5){$1$}
\put(1.5,-.5){$2$} \put(6,-.5){$1$} \put(7.5,-.5){$2$}
\end{picture}
\end{array}
\label{2.2}%
\end{equation}
and crossing
\begin{gather}
S_{12}(\theta_{1}-\theta_{2})=\mathbf{C}^{2\bar{2}}\,S_{\bar{2}1}(\theta
_{2}+i\pi-\theta_{1})\,\mathbf{C}_{\bar{2}2}=\mathbf{C}^{1\bar{1}}%
\,S_{2\bar{1}}(\theta_{2}-(\theta_{1}-i\pi))\,\mathbf{C}^{\bar{1}1}%
\label{2.4}\\%
\begin{array}
[c]{c}%
\unitlength3mm\begin{picture}(4,5) \put(0,1){\line(1,1){4}}
\put(4,1){\line(-1,1){4}} \put(0,-.5){$1$} \put(3.7,-.5){$2$}
\end{picture}
\end{array}
~~~~~~=~~~~~~%
\begin{array}
[c]{c}%
\unitlength3mm\begin{picture}(6,5) \put(1,1){\line(1,1){4}}
\put(4,1){\line(-1,2){2}} \put(1,5){\oval(2,8)[lb]}
\put(5,1){\oval(2,8)[tr]} \put(3.5,-.5){$1$} \put(5.7,-.5){$2$}
\end{picture}
\end{array}
~~~~~~=~~~~~~%
\begin{array}
[c]{c}%
\unitlength3mm\begin{picture}(6,5) \put(2,1){\line(1,2){2}}
\put(5,1){\line(-1,1){4}} \put(1,1){\oval(2,8)[lt]}
\put(5,5){\oval(2,8)[br]} \put(0,-.5){$1$} \put(2,-.5){$2$}
\end{picture}
\end{array}
\nonumber
\end{gather}
where $\mathbf{C}^{1\bar{1}}$ and $\mathbf{C}_{1\bar{1}}$ are
charge conjugation matrices. We have introduced the following
graphical rule, that a line changing the ``time direction'' also
interchanges particles and anti-particles and changes the rapidity
as $\theta\rightarrow\theta\pm i\pi$
\begin{equation}
\mathbf{C}_{\alpha\bar{\beta}}=%
\begin{array}
[c]{c}%
\unitlength4mm\begin{picture}(5.5,3) \put(2,1){\oval(2,4)[t]}
\put(0,1){$\theta$} \put(3.3,1){$\theta-i\pi$}
\put(.7,0){$\alpha$} \put(2.7,-.1){$\bar\beta$} \end{picture}
\end{array}
,\qquad \mathbf{C}^{\alpha\bar{\beta}}=%
\begin{array}
[c]{c}%
\unitlength4mm\begin{picture}(5.5,3) \put(2,2){\oval(2,4)[b]}
\put(0,1){$\theta$} \put(3.3,1){$\theta+i\pi$}
\put(.7,2.2){$\alpha$} \put(2.7,2.2){$\bar\beta$} \end{picture}
\end{array}
. \label{2.6}%
\end{equation}

\paragraph{Bound states:}

Let $\gamma$ be a bound state of particles $\alpha$ and $\beta$
with mass
\[
m_{\gamma}=\sqrt{m_{\alpha}^{2}+m_{\beta}^{2}+2m_{\alpha}m_{\beta}\cos\eta
}\,,\qquad (0<\eta<\pi).
\]
Then the 2 particle S-matrix has a pole such that
\begin{align}
i\operatorname*{Res}_{\theta=i\eta}S_{\alpha\beta}^{\beta^{\prime}%
\alpha^{\prime}}(\theta)  &  =\Gamma_{\gamma}^{\beta^{\prime}\alpha^{\prime}%
}\,\Gamma_{\alpha\beta}^{\gamma}\label{2.8}\\[0.2in]
i\operatorname*{Res}_{\theta=i\eta}%
\begin{array}
[c]{c}%
\unitlength2.7mm\begin{picture}(5,6) \thicklines
\put(0,1){\line(1,1){4}} \put(2,3){\makebox(0,0){$\bullet$}}
\put(4,1){\line(-1,1){4}} \put(0,-.3){$\alpha$}
\put(3.5,-.3){$\beta$} \put(3.5,5.5){$\alpha'$}
\put(0,5.5){$\beta'$} \end{picture}
\end{array}
~~  &  =~~~%
\begin{array}
[c]{c}%
\unitlength2mm\begin{picture}(5,8) \thicklines
\put(0,1){\line(1,1){2}} \put(2,3){\makebox(0,0){$\bullet$}}
\put(4,1){\line(-1,1){2}} \put(2,3){\line(0,1){2}}
\put(2,5){\makebox(0,0){$\bullet$}} \put(2,5){\line(1,1){2}}
\put(2,5){\line(-1,1){2}} \put(0,-.5){$\alpha$}
\put(3.5,-.5){$\beta$} \put(2.5,3.7){$\gamma$}
\put(3.5,7.7){$\alpha'$} \put(0,7.7){$\beta'$} \end{picture}
\end{array}
\nonumber
\end{align}
where $\eta$ is called the fusion angle and
$\Gamma_{\alpha\beta}^{\gamma}$ is the `bound state intertwiner'
\cite{K1,BK}. The \textbf{bound state S-matrix, }that is the
scattering matrix of the bound state (12) with a particle 3, is
obtained by the \textbf{``bootstrap equation''} \cite{K1}
\begin{align*}
S_{(12)3}(\theta_{(12)3})\,\Gamma_{12}^{(12)}  &  =\Gamma_{12}^{(12)}%
\,S_{13}(\theta_{13})S_{23}(\theta_{23})\,\\%
\begin{array}
[c]{c}%
\unitlength3mm\begin{picture}(5,5) \thicklines
\put(2,2){\line(1,1){2}} \put(4.5,1.5){\line(-1,1){3}}
\put(2,2){\line(-4,-1){2}} \put(2,0){\line(0,1){2}}
\put(0,.4){$1$} \put(1,0){$2$} \put(3.5,.7){$3$}
\put(4.2,4){$(12)$} \put(2,2){\makebox(0,0){$\bullet$}}
\end{picture}
\end{array}
~~~  &  =~~~%
\begin{array}
[c]{c}%
\unitlength3mm\begin{picture}(5,5) \thicklines
\put(3,3){\line(1,1){1.2}} \put(4,0){\line(-1,1){4}}
\put(0,2){\line(3,1){3}} \put(2,0){\line(1,3){1}} \put(0,.8){$1$}
\put(1,0){$2$} \put(4,.4){$3$} \put(4.4,4){$(12)$}
\put(3,3){\makebox(0,0){$\bullet$}} \end{picture}
\end{array}
\end{align*}
where we use the usual short hand notation of matrices acting in
the spaces corresponding to the particles 1, 2, 3 and (12).

\paragraph{\textbf{Examples}}
of integrable models in 1+1-dimensions (which we will consider in
this review) are the \textbf{sinh-Gordon} model def\/ined by the
classical f\/ield equation
\[
\ddot{\varphi}(t,x)-\varphi^{\prime\prime}(t,x)+\frac{\alpha}{\beta}\sinh
\beta\varphi(t,x)=0
\]
and the $\boldsymbol{SU(N)}$ \textbf{Gross--Neveu} model described by
the Lagrangian
\[
\mathcal{L}=\bar{\psi}\,i\gamma\partial\,\psi+\frac{g^{2}}{2}\left(
(\bar{\psi}\psi)^{2}-(\bar{\psi}\gamma^{5}\psi)^{2}\right)  ,
\]
where the Fermi f\/ields form an $SU(N)$ multiplet.

Further integrable quantum f\/ield theories are: scaling
$Z_{N}$-Ising models, nonlinear $\sigma$-models, $O(N)
$Gross--Neveu models, Toda models etc.

\subsection*{The S-matrix}

\paragraph{The sinh-Gordon S-matrix}
is given by the analytic continuation $\beta\rightarrow i\beta$ of
the sin-Gordon breather S-matrix \cite{KT}
\begin{equation}
S(\theta)=\frac{\sinh\theta+i\sin\pi\nu}{\sinh\theta-i\sin\pi\nu}%
\qquad\text{with} \quad
-1\leq\nu=\frac{-\beta^{2}}{8\pi+\beta^{2}}\leq0.
\label{2.10}%
\end{equation}
The model has the self-dual point at%
\[
\nu=-\,\frac{1}{2}\qquad \mathrm{or}\qquad \beta^{2}=4\pi.
\]

\paragraph{The $\boldsymbol{SU(N)}$ S-matrix:}

All solutions of $U(N)$-invariant S-matrix satisfying unitarity,
crossing and the Yang--Baxter equation have been obtained in
\cite{BKKW}. Following \cite{KuS,KS}, we adopt the view that in
the $SU(N)$ Gross--Neveu model, the anti-particles are bound
states of $N-1$ particles. This implies that there is no particle
anti-particle backward scattering and that the $SU(N)$ S-matrix
should be given by solution II of \cite{BKKW}. The scattering of
the fundamental particles which form a multiplet corresponding to
the vector representation of $SU(N)$ is (see \cite{KKS,BW,ABW} and
\cite{Be} for $N=2$)
\begin{equation}
S_{\alpha\beta}^{\delta\gamma}(\theta)=%
\begin{array}
[c]{c}%
\unitlength3mm\begin{picture}(6,6) \put(1,1){\line(1,1){4}}
\put(1,1){\vector(1,1){1.2}} \put(3,3){\vector(1,1){1.2}}
\put(5,1){\line(-1,1){4}} \put(5,1){\vector(-1,1){1.2}}
\put(3,3){\vector(-1,1){1.2}} \put(.5,.1){$\scriptstyle\alpha$}
\put(5,.1){$\scriptstyle\beta$} \put(5,5.3){$\scriptstyle\gamma$}
\put(.5,5.3){$\scriptstyle\delta$} \put(1.8,1){$\scriptstyle p_1$}
\put(3.5,1){$\scriptstyle p_2$} \put(3.5,4.7){$\scriptstyle p_3$}
\put(1.8,4.7){$\scriptstyle p_4$} \end{picture}
\end{array}
=\delta_{\alpha\gamma}\delta_{\beta\delta}\,b(\theta)+\delta_{\alpha\delta
}\delta_{\beta\gamma}\,c(\theta) \label{2.12}%
\end{equation}
where due to Yang--Baxter $c(\theta)=-\frac{2\pi
i}{N\theta}\,b(\theta)$ holds and the highest weight amplitude is
given as
\begin{equation}
a(\theta)=b(\theta)+c(\theta)=-\frac{\Gamma\left(
1-\frac{\theta}{2\pi
i}\right)  \Gamma\left(  1-\frac{1}{N}+\frac{\theta}{2\pi i}\right)  }%
{\Gamma\left(  1+\frac{\theta}{2\pi i}\right)  \Gamma\left(  1-\frac{1}%
{N}-\frac{\theta}{2\pi i}\right)  }\,. \label{2.14}%
\end{equation}
There is a bound state pole at $\theta=i\eta=2\pi i/N$ in the
antisymmetric tensor sector which agrees with Swieca's \cite{KuS}
picture that the bound state of $N-1$ particles is to be
identif\/ied with the anti-particle. Similar as in the scaling
$Z(N)$-Ising and $A(N-1)$-Toda models \cite{BK04,BFK} this will be
used to construct the form factors in $SU(N)$ model \cite{BFK1}.

\section{Form factors}\label{s3}

For a local operator $\mathcal{O}(x)$ the generalized form factors
\cite{KW} are def\/ined as
\begin{equation}
\,F_{\alpha_{1}\dots\alpha_{n}}^{\mathcal{O}}\left(
\theta_{1},\dots ,\theta_{n}\right)
=\langle\,0\,|\,\mathcal{O}(0)\,|\,p_{1},\dots
,p_{n}\,\rangle_{\alpha_{1}\dots\alpha_{n}}^{\rm in}\label{1.8}%
\end{equation}
for $\theta_{1}>\dots>\theta_{n}$. For other orders of the
rapidities they are def\/ined by analytic continuation. The index
$\alpha_{i}$\thinspace denotes the type of the particle with
momentum $p_{i}$. We also use the short notations
$F_{\underline{\alpha}}^{\mathcal{O}}(\underline{\theta})$ or
$F_{1\dots n}^{\mathcal{O}}(\underline{\theta})$\footnote{The
later means the co-vector
in a tensor product space with the components $F_{\underline{\alpha}%
}^{\mathcal{O}}$.}.

For the $SU(N)$ Gross--Neveu model $\alpha$ denotes the types of
particles belonging to all fundamental representations of $SU(N)$
with dimension $\binom{N}{r},~r=1,\dots,N-1$. In most formulae we
restrict $\alpha=1,\dots,N$ to the multiplet of the vector
representation. Similar as for the S-matrix, `maximal analyticity'
for generalized form factors means again that they are
meromorphic and all poles in the `physical strips' $0\leq\operatorname{Im}%
\theta_{i}\leq\pi$ have a physical interpretation. Together with
the usual LSZ-assumptions \cite{LSZ} of local quantum f\/ield
theory the following form factor equations can be derived:

\begin{itemize}\itemsep=0pt
\item[$(i)$] The Watson's equations describe the symmetry property
under the permutation of both, the variables
$\theta_{i},\theta_{j}$ and the spaces $i,j=i+1$ at the same time
\begin{equation}
F_{\dots
ij\dots}^{\mathcal{O}}(\dots,\theta_{i},\theta_{j},\dots)=F_{\dots
ji\dots}^{\mathcal{O}}(\dots,\theta_{j},\theta_{i},\dots)\,S_{ij}(\theta_{ij})
\label{3.2}%
\end{equation}
for all possible arrangements of the $\theta$'s.

\item[$(ii)$] The crossing relation which implies a periodicity
property under the cyclic permutation of the rapidity variables
and spaces
\begin{gather}
^{\text{out,}\bar{1}}\langle\,p_{1}\,|\,\mathcal{O}(0)\,|\,p_{2},\dots
,p_{n}\,\rangle_{2\dots n}^{\text{in,conn.}}\nonumber\\
\qquad{}=\mathbf{C}^{\bar{1}1}\sigma_{1}^{\mathcal{O}}F_{1\ldots n}^{\mathcal{O}%
}(\theta_{1}+i\pi,\theta_{2},\dots,\theta_{n})=F_{2\ldots n1}^{\mathcal{O}%
}(\theta_{2},\dots,\theta_{n},\theta_{1}-i\pi)\mathbf{C}^{1\bar{1}}
\label{3.4}%
\end{gather}
where $\sigma_{\alpha}^{\mathcal{O}}$ takes into account the
statistics of the particle $\alpha$ with respect to $\mathcal{O}$
(e.g., $\sigma_{\alpha }^{\mathcal{O}}=-1$ if $\alpha$ and
$\mathcal{O}$ are both fermionic, these numbers can be more
general for anyonic or order and disorder f\/ields, see
\cite{BFK}). For the charge conjugation matrix
$\mathbf{C}^{\bar{1}1}$ we refer to (\ref{2.6}).

\item[$(iii)$] There are poles determined by one-particle states
in each sub-channel given by a subset of particles of the state in
(\ref{1.8}).

In particular the function $F_{\underline{\alpha}}^{\mathcal{O}%
}({\underline{\theta}})$ has a pole at $\theta_{12}=i\pi$ such
that
\begin{equation}
\operatorname*{Res}_{\theta_{12}=i\pi}F_{1\dots
n}^{\mathcal{O}}(\theta
_{1},\dots,\theta_{n})=2i\mathbf{C}_{12}\,F_{3\dots n}^{\mathcal{O}}%
(\theta_{3},\dots,\theta_{n})\left(  \mathbf{1}-\sigma_{2}^{\mathcal{O}}%
S_{2n}\cdots S_{23}\right)  . \label{3.6}%
\end{equation}

\item[$(iv)$] If there are also bound states in the model the
function
$F_{\underline{\alpha}}^{\mathcal{O}}({\underline{\theta}})$ has
additional poles. If for instance the particles 1 and 2 form a
bound state (12), there is a pole at $\theta_{12}=i\eta$
$(0<\eta<\pi)$ such that
\begin{equation}
\operatorname*{Res}_{\theta_{12}=\eta}F_{12\dots
n}^{\mathcal{O}}(\theta
_{1},\theta_{2},\dots,\theta_{n})=F_{(12)\dots n}^{\mathcal{O}}%
(\theta_{(12)},\dots,\theta_{n}) \sqrt{2}\Gamma_{12}^{(12)} \label{3.8}%
\end{equation}
where the bound state intertwiner $\Gamma_{12}^{(12)}$ is
def\/ined by (\ref{2.8}) (see \cite{K1,BK}).

\item[$(v)$] Naturally, since we are dealing with relativistic
quantum f\/ield theories we f\/inally have
\begin{equation}
F_{1\dots
n}^{\mathcal{O}}(\theta_{1}+\mu,\dots,\theta_{n}+\mu)=e^{s\mu
} F_{1\dots n}^{\mathcal{O}}(\theta_{1},\dots,\theta_{n}) \label{3.10}%
\end{equation}
if the local operator transforms under Lorentz transformations as
$F^{\mathcal{O}}\rightarrow e^{s\mu}F^{\mathcal{O}}$ where $s$ is
the ``spin'' of $\mathcal{O}$.
\end{itemize}

The properties $(i)$--$(iv)$ may be depicted as
\[%
\begin{array}
[c]{rrcl}%
(i) &
\begin{array}
[c]{c}%
\unitlength3.2mm\begin{picture}(7,3) \put(3.5,2){\oval(7,2)}
\put(3.5,2){\makebox(0,0){${\cal O}$}} \put(1,0){\line(0,1){1}}
\put(3,0){\line(0,1){1}} \put(4,0){\line(0,1){1}}
\put(6,0){\line(0,1){1}} \put(1.4,.5){$\dots$}
\put(4.4,.5){$\dots$} \end{picture}
\end{array}
& = &
\begin{array}
[c]{c}%
\unitlength3.2mm\begin{picture}(7,4) \put(3.5,3){\oval(7,2)}
\put(3.5,3){\makebox(0,0){${\cal O}$}} \put(1,0){\line(0,1){2}}
\put(3,0){\line(1,2){1}} \put(4,0){\line(-1,2){1}}
\put(6,0){\line(0,1){2}} \put(1.4,1){$\dots$} \put(4.4,1){$\dots$}
\end{picture}
\end{array}
\\
(ii) & ~~~~~~%
\begin{array}
[c]{c}%
\unitlength3.2mm\begin{picture}(5,4)(2,0) \put(2,0){\line(0,1){1}}
\put(3.5,3){\line(0,1){1}} \put(3.5,2){\oval(5,2)}
\put(3.5,2){\makebox(0,0){${\cal O}$}} \put(5,0){\line(0,1){1}}
\put(6,3){\scriptsize conn.} \put(2.8,.3){$\dots$} \end{picture}
\end{array}
~~=~~\sigma^{\mathcal{O}}%
\begin{array}
[c]{c}%
\unitlength3.2mm\begin{picture}(6,4) \put(0,1){\line(0,1){3}} \put
(1,1){\oval(2,2)[b]} \put(3.5,2){\oval(5,2)}
\put(3.5,2){\makebox(0,0){${\cal O}$}} \put(5,0){\line(0,1){1}}
\put(2.6,.3){$\dots$} \end{picture}
\end{array}
~~ & = &
\begin{array}
[c]{c}%
\unitlength3.2mm\begin{picture}(7,4) \put(7,1){\line(0,1){3}} \put
(6,1){\oval(2,2)[b]} \put(3.5,2){\oval(5,2)}
\put(3.5,2){\makebox(0,0){${\cal O}$}} \put(2,0){\line(0,1){1}}
\put(2.6,.3){$\dots$} \end{picture}
\end{array}
\\
(iii) & \dfrac{1}{2i}\,\operatorname*{Res}\limits_{\theta_{12}=i\pi}~~~%
\begin{array}
[c]{c}%
\unitlength3.2mm\begin{picture}(6,4) \put(3,2){\oval(6,2)}
\put(3,2){\makebox(0,0){${\cal O}$}} \put(1,0){\line(0,1){1}}
\put(2,0){\line(0,1){1}} \put(3,0){\line(0,1){1}}
\put(5,0){\line(0,1){1}} \put(3.4,.5){$\dots$} \end{picture}
\end{array}
& = &
\begin{array}
[c]{c}%
\unitlength3.2mm\begin{picture}(5,4) \put(.5,0){\oval(1,2)[t]}
\put(3,2){\oval(4,2)} \put(3,2){\makebox(0,0){${\cal O}$}}
\put(2,0){\line(0,1){1}} \put(4,0){\line(0,1){1}}
\put(2.4,.5){$\dots$} \end{picture}
\end{array}
~~-\sigma^{\mathcal{O}}~~%
\begin{array}
[c]{c}%
\unitlength3.2mm\begin{picture}(6,5) \put(0,0){\line(0,1){3}}
\put(3,3){\oval(6,4)[t]} \put(3,3){\oval(6,4)[br]}
\put(3,0){\oval(4,2)[tl]} \put(3,3){\oval(4,2)}
\put(3,3){\makebox(0,0){${\cal O}$}} \put(2,0){\line(0,1){2}}
\put(4,0){\line(0,1){2}} \put(2.4,1.5){$\dots$} \end{picture}
\end{array}
\\
(iv) & \dfrac{1}{\sqrt{2}}\,\operatorname*{Res}\limits_{\theta_{12}=\eta}~%
\begin{array}
[c]{c}%
\unitlength3.2mm\begin{picture}(5,3) \put(2.5,2){\oval(5,2)}
\put(2.5,2){\makebox(0,0){${\cal O}$}} \put(1,0){\line(0,1){1}}
\put(2,0){\line(0,1){1}} \put(4,0){\line(0,1){1}}
\put(2.4,.5){$\dots$} \end{picture}
\end{array}
& = &
\begin{array}
[c]{c}%
\unitlength3.2mm\begin{picture}(5,4) \put(2.5,3){\oval(5,2)}
\put(2.5,3){\makebox(0,0){${\cal O}$}} \put(1.5,0){\oval(1,2)[t]}
\put(1.5,1){\line(0,1){1}} \put(4,0){\line(0,1){2}}
\put(2.4,1){$\dots$} \end{picture}
\end{array}
\end{array}
\]

These equations have been proposed by Smirnov \cite{Sm} as
generalizations of equations derived in the original articles
\cite{KW,BKW,K2}. They have been proven \cite{BFKZ} by means of
the LSZ-assumptions and `maximal analyticity'.

We will now provide a constructive and systematic way of how to
solve the equations $(i)$--$(v)$ for the co-vector valued function
$F_{1\dots n}^{\mathcal{O}}$ once the scattering matrix is given.

\subsection{Two-particle form factors}

For the two-particle form factors the form factor equations are
easily understood. The usual assumptions of local quantum f\/ield
theory yield
\[
\langle\,0\,|\,\mathcal{O}(0)\,|\,p_{1},p_{2}\rangle^{\rm
in/out}=F\left( (p_{1}+p_{2})^{2}\pm i\varepsilon\right)  =
F\left(  \pm\theta_{12}\right)
\]
where the rapidity dif\/ference is def\/ined by
$p_{1}p_{2}=m^{2}\cosh\theta_{12}$. For integrable theories one
has particle number conservation which implies (for any eigenstate
of the two-particle S-matrix)
\[
\langle\,0\,|\,\mathcal{O}(0)\,|\,p_{1},p_{2}\rangle^{\rm
in}=\langle \,0\,|\,\mathcal{O}(0)\,|\,p_{2},p_{1}\rangle^{\rm
out} S\left(  \theta _{12}\right)  .
\]
Crossing (\ref{3.4}) means
\[
\langle\,p_{1}\,|\,\mathcal{O}(0)\,|\,p_{2}\rangle=F\left(
i\pi-\theta _{12}\right)
\]
where for one-particle states the in- and out-states coincide.
Therefore Watson's equations follow
\begin{gather}%
F\left(  \theta\right)  =F\left(  -\theta\right)  S\left(  \theta\right),\nonumber\\
F\left(  i\pi-\theta\right)  =F\left(  i\pi+\theta\right)  .
\label{3.12}%
\end{gather}
For general theories Watson's \cite{Wa} equations only hold below
the particle production thresholds. However, for integrable
theories there is no particle production and therefore they hold
for all complex values of $\theta$. It has been shown \cite{KW}
that these equations together with ``maximal analyticity'' have a
unique solution.

As an example we write the sinh-Gordon \cite{KW} and the (highest
weight) $SU(N)$ form factor function \cite{BFK1}
\begin{gather}
F^{SHG}(\theta)   =\exp\int_{0}^{\infty}\frac{dt}{t\sinh
t}\,\left(
\frac{\cosh(\frac{1}{2}+\nu)t}{\cosh\frac{1}{2}t}-1\right)  \cosh
t\left(
1-\frac{\theta}{i\pi}\right), \label{3.14}\\
F^{SU(N)}\left(  \theta\right)
=c\exp\int\limits_{0}^{\infty}\frac
{dt}{t\sinh^{2}t}e^{\frac{t}{N}}\sinh t\left(
1-\frac{1}{N}\right)  \left( 1-\cosh t\left(
1-\frac{\theta}{i\pi}\right)  \right)
\end{gather}
which are the minimal solution of (\ref{3.12}) with $S^{SHG}\left(
\theta\right)  $ as given by (\ref{2.10}) and $S^{SU(N)}\left(
\theta\right) =a\left(  \theta\right)  $ as given by (\ref{2.14}),
respectively.

\subsection{The general form factor formula}

As usual \cite{KW} we split of\/f the minimal part and write the
form factor for
n particles as%
\begin{equation}
F_{\alpha_{1}\dots\alpha_{n}}^{\mathcal{O}}(\theta_{1},\dots,\theta
_{n})=K_{\alpha_{1}\dots\alpha_{n}}^{\mathcal{O}}(\underline{\theta}%
)\prod\limits_{1\leq i<j\leq n}F(\theta_{ij}).\label{3.16}%
\end{equation}
By means of the following {``of\/f-shell Bethe ansatz''} for the
(co-vector valued) K-function%
\begin{equation}
\fbox{$\rule[-0.18in]{0in}{0.45in}~K_{\alpha_{1}\dots\alpha_{n}}^{\mathcal{O}%
}(\underline{\theta})=\displaystyle\int_{\mathcal{C}_{\underline{\theta}}%
}dz_{1}\cdots\displaystyle\int_{\mathcal{C}_{\underline{\theta}}}%
dz_{m}\,h(\underline{\theta},\underline{z})\,p^{\mathcal{O}}(\underline
{\theta},\underline{z})\,\Psi_{\alpha_{1}\dots\alpha_{n}}(\underline{\theta
},\underline{z})~$}\label{3.18}%
\end{equation}
we transform the complicated form factor equations $(i)-(v)$ into
simple ones for the p-functions which are scalar and polynomials
in $e^{\pm z_{i}}$. The \textbf{``of\/f-shell Bethe ansatz''}{
state }$\Psi_{\alpha
_{1}\dots\alpha_{n}}(\underline{\theta},\underline{z})$ is
obtained as a
product of S-matrix elements and the integration contour $\mathcal{C}%
_{\underline{\theta}}$ depends on the model (see below for the
$SU(N)$-model). The scalar functions
\begin{gather}
h(\underline{\theta},\underline{z})   =\prod_{i=1}^{n}\prod_{j=1}^{m}%
\phi(\theta_{i}-z_{j})\prod_{1\leq i<j\leq m}\tau(z_{i}-z_{j}),\label{3.20}%
\\
\tau(z)  =\frac{1}{\phi(z)\phi(-z)}\nonumber
\end{gather}
depend on $S(\theta)$ only (see (\ref{3.21}) below), i.e.~on the
S-matrix, whereas the $p$-function
$p^{\mathcal{O}}(\underline{\theta},\underline{z})$ depends on the
operator.

\subsubsection*{The $\boldsymbol{SU(N)}$ form factors}

The form factors for n fundamental particles (of the vector
representation of $SU(N)$) are given by (\ref{3.16})--(\ref{3.20})
where the \textbf{``nested Bethe ansatz''} is needed. This means
that
$\Psi${ is of the form}%
\begin{equation}
\Psi_{\alpha_{1}\dots\alpha_{n}}(\underline{\theta},\underline{z}%
)=L_{\beta_{1}\dots\beta_{m}}(\underline{z})\Phi_{\alpha_{1}\dots\alpha_{n}%
}^{\beta_{1}\dots\beta_{m}}(\underline{\theta},\underline{z}) \label{3.29}%
\end{equation}
where the indices $\alpha_{i}$ take the values $\alpha=1,\dots,N$
and the summations run over $\beta_{i}=2,\dots,N$. The
\textbf{``of\/f-shell Bethe ansatz''} state
$\Phi_{\alpha_{1}\dots\alpha_{n}}^{\beta
_{1}\dots\beta_{m}}(\underline{\theta},\underline{z})$ is obtained
using the techniques of the algebraic Bethe ansatz as follows.

We consider a state with $n$ particles and def\/ine the monodromy
matrix
\begin{equation}
T_{1\dots
n,0}(\underline{\theta},\theta_{0})=S_{10}(\theta_{10})\,\cdots
S_{n0}(\theta_{n0})=%
\begin{array}
[c]{c}%
\unitlength3mm\begin{picture}(9,4.5) \put(0,2){\line(1,0){9}}
\put(2,0){\line(0,1){4}} \put(7,0){\line(0,1){4}} \put(1,0){$1$}
\put(5.8,0){$ n$} \put(8.2,.7){$ 0$} \put(3.6,3){$\dots$}
\end{picture}
\end{array}
\nonumber
\end{equation}
as a matrix acting in the tensor product of the ``quantum space'',
a space of $n$ particles (with respect to their quantum numbers)
$V^{1\dots n}=V^{1}\otimes\cdots\otimes V^{n}$ and the ``auxiliary
space'' $V^{0}$. All vector spaces $V^{i}$ are isomorphic to a
space $V$ whose basis vectors label all kinds of particles. Here
we consider $V\cong\mathbb{C}^{N}$ as the space of the vector
representation of $SU(N)$. The Yang--Baxter algebra relation for
the S-matrix yields
\begin{gather}
T_{1\dots n,a}(\underline{\theta},\theta_{a})\,T_{1\dots
n,b}(\underline
{\theta},\theta_{b})\,S_{ab}(\theta_{a}-\theta_{b})=S_{ab}(\theta_{a}%
-\theta_{b})\,T_{1\dots
n,b}(\underline{\theta},\theta_{b})\,T_{1\dots
n,a}(\underline{\theta},\theta_{a})\label{1.32}\\
\unitlength4mm\begin{picture}(19,5) \put(.3,0){$1$}
\put(4.3,0){$n$} \put(0,2.3){$b$} \put(0,4.3){$a$}
\put(7.5,2.3){$b$} \put(7.5,.5){$a$} \put(2.5,3){$\dots$}
\put(1,0){\line(0,1){5}} \put(0,2){\line(1,0){8}}
\put(5,0){\line(0,1){5}} \put(0,2){\oval(14,4)[rt]}
\put(8,2){\oval(2,4)[lb]} \put(9.3,2.3){$=$} \put(13.3,0){$1$}
\put(17.3,0){$n$} \put(11,2){$b$} \put(11,4.2){$a$}
\put(18.5,3.4){$b$} \put(18.5,1.4){$a$} \put(15.5,2){$\dots$}
\put(14,0){\line(0,1){5}} \put(18,0){\line(0,1){5}}
\put(11,3){\line(1,0){8}} \put(19,3){\oval(14,4)[lb]}
\put(11,3){\oval(2,4)[rt]} \end{picture}\nonumber
\end{gather}
which in turn implies the basic algebraic properties of the
sub-matrices $A$, $B$, $C$, $D$ with respect to the auxiliary
space def\/ined by
\begin{equation}
T_{1\dots n,0}(\underline{\theta},z)\equiv\left(
\begin{array}
[c]{cc}%
A_{1\dots n}(\underline{\theta},z) & B_{1\dots
n,\beta}(\underline{\theta
},z)\\
C_{1\dots n}^{\beta}(\underline{\theta},z) & D_{1\dots
n,\beta}^{\beta ^{\prime}}(\underline{\theta},z)
\end{array}
\,\right)  ,\qquad 2\leq\beta,\beta^{\prime}\leq N. \label{1.34}%
\end{equation}
The basic Bethe ansatz co-vectors $\Phi$ of equation (\ref{3.29})
are obtained by an application of the operators $C$ to a ``pseudo-vacuum'' state%
\begin{equation}
\Phi_{1\dots n}^{\underline{\beta}}(\underline{\theta},\underline{z}%
)=\Omega_{1\dots n}C_{1\dots
n}^{\beta_{m}}(\underline{\theta},z_{m})\cdots
C_{1\dots n}^{\beta_{1}}(\underline{\theta},z_{1}). \label{1.38}%
\end{equation}
This may be depicted as%
\[%
\begin{array}
[c]{r}%
\Phi_{\underline{\alpha}}^{\underline{\beta}}(\underline{\theta},\underline
{z})=%
\begin{array}
[c]{c}%
\unitlength4mm\begin{picture}(12,8) \put(9,5){\oval(14,2)[lb]}
\put(9,5){\oval(18,6)[lb]} \put(4,1){\line(0,1){4}}
\put(8,1){\line(0,1){4}} \put(3.7,0){$\alpha_1$}
\put(7.8,0){$\alpha_n$} \put(-.2,5.4){$\beta_1$}
\put(1.8,5.4){$\beta_m$} \put(3.8,5.4){$1$} \put(7.8,5.4){$1$}
\put(9.2,1.8){$1$} \put(9.2,3.8){$1$} \put(4.3,1){$\theta_1$}
\put(8.3,1){$\theta_{n}$} \put(5.5,2.342){$z_1$}
\put(5.5,4.3){$z_m$} \put(.6,4.3){$\dots$} \put(5.8,1.3){$\dots$}
\put(8.6,2.6){$\vdots$} \end{picture}
\end{array}
\text{with }
\begin{array}
[c]{l}%
2\leq\beta_{i}\leq N,\\
1\leq\alpha_{i}\leq N.
\end{array}
\end{array}
\]
It means that
$\Phi_{\underline{\alpha}}^{\underline{\beta}}(\underline
{\theta},\underline{z})$ is a product of S-matrix elements as
given by the picture where at all crossing points of lines there
is an S-matrix (\ref{2.12}) and the sum over all indices of
internal lines is to be taken. The ``pseudo-vacuum'' is the
highest weight co-vector (with weight $w=(n,0,\dots,0)$)
\[
\Omega_{1\dots n}=e(1)\otimes\cdots\otimes e(1)
\]
where the unit vectors $e(\alpha)$ $(\alpha=1,\dots,N)$ correspond
to the particle of type $\alpha$ which belong to the vector
representation of $SU(N)$. The pseudo-vacuum\ basic vector
satisf\/ies
\begin{gather}
\Omega_{1\dots n} B_{1\dots n}^{\beta}(\underline{\theta},z)  =  0,\nonumber\\
\Omega_{1\dots n}\,A_{1\dots n}(\underline{\theta},z)  =  \prod
\limits_{i=1}^{n}a(\theta_{i}-z)\Omega_{1\dots n},\label{1.40}\\
\Omega_{1\dots n}\,D_{1\dots
n,\beta}^{\beta^{\prime}}(\underline{\theta},z)
=  \delta_{\beta}^{\beta^{\prime}}\prod\limits_{i=1}^{n}b(\theta_{i}%
-z)\Omega_{1\dots n}.\nonumber
\end{gather}
The amplitudes of the scattering matrices are given by equation
(\ref{2.12}). The technique of the nested Bethe ansatz means that
for the co-vector valued function
$L_{\beta_{1}\dots\beta_{m}}(\underline{z})$ in (\ref{3.29}) one
makes the second level Bethe ansatz. This ansatz is of the same
form as (\ref{3.18}) only that the range of the indices is reduced
by 1. Iterating this nesting procedure one f\/inally arrives at a
scalar function. The integration contour
$\mathcal{C}_{\underline{\theta}}$ depends on the
$\underline{\theta}$ and is depicted in Fig.~\ref{f2}.
\begin{figure}[tbh]
\begin{center}
\unitlength4mm\begin{picture}(27,13) \thicklines \put(1,0){
\put(0,0){$\bullet~\theta_n-2\pi i$}
\put(.19,5){\circle{.3}~$\theta_n-2\pi i\frac1N$}
\put(0,6){$\bullet~~\theta_n$} \put(.2,6.2){\oval(1,1)}
\put(.5,6.68){\vector(1,0){0}} \put(0,11){$\bullet~\theta_n+2\pi
i(1-\frac1N)$} } \put(8,6){\dots} \put(12,0){
\put(0,0){$\bullet~\theta_2-2\pi i$}
\put(.19,5){\circle{.3}~$\theta_2-2\pi i\frac1N$}
\put(0,6){$\bullet~~\theta_2$} \put(.2,6.2){\oval(1,1)}
\put(.5,6.68){\vector(1,0){0}} \put(0,11){$\bullet~\theta_2+2\pi
i(1-\frac1N)$} } \put(20,1){ \put(0,0){$\bullet~\theta_1-2\pi i$}
\put(.19,5){\circle{.3}~$\theta_1-2\pi i\frac1N$}
\put(0,6){$\bullet~~\theta_1$} \put(.2,6.2){\oval(1,1)}
\put(.5,6.68){\vector(1,0){0}} \put(0,11){$\bullet~\theta_1+2\pi
i(1-\frac1N)$} } \put(9,2.5){\vector(1,0){0}}
\put(0,3){\oval(34,1)[br]} \put(27,3){\oval(20,1)[tl]}
\end{picture}
\end{center}
\caption{\emph{The integration contour
}$C_{\underline{\theta}}$\emph{. The
bullets belong to poles of the integrand resulting from }$a(\theta_{i}%
-z_{j})\,\phi(\theta_{i}-z_{j})$\emph{ and the small open circles
belong to
poles originating from }$b(\theta_{i}-z_{j})$\emph{ and }$c(\theta_{i}-z_{j}%
)$\emph{. }}%
\label{f2}%
\end{figure}

Swieca's picture is that the bound state of $N-1$ fundamental
particles is to be identif\/ied with the anti-particle and lead
together with the form factor recursion relations $(iii)+(iv)$ to
the equation for the function $\phi(z)$ (see \cite{BFK,BFK1})
\begin{equation}
\prod_{k=0}^{N-2}\phi\left(  z+ki\eta\right)
\prod_{k=0}^{N-1}F\left(
z+ki\eta\right)  =1,\qquad \eta=\frac{2\pi}{N} \label{3.21}%
\end{equation}
with the solution
\[
\fbox{ $\phi(\theta)=\Gamma\left(  \dfrac{\theta}{2\pi i}\right)
\Gamma\left(  1-\dfrac{1}{N}-\dfrac{\theta}{2\pi i}\right).$}
\]
In \cite{BFK1} it is shown that the form factors given by
(\ref{3.16}) and (\ref{3.18}) satisfy the form factor equations
$(i)$--$(v)$ if some simple equations for the p-function are
satisf\/ied. We note that form factors of this model were also
calculated in \cite{Sm,NT,Ta} using other techniques.

\subsubsection*{Locality}

Let us discuss a further property of the form factors which is
directly related to the nature of the operators, that is
\textit{locality}. In fact, this property touches the very heart
and most central concepts of relativistic quantum f\/ield theory,
like Einstein causality and Poincar\'{e} covariance, which are
captured in local f\/ield equations and commutation relations.

So far we have presented a formulation of a quantum f\/ield
theory, which starts from a particle picture. It is basically
possible to obtain the particle picture from the f\/ield
formulation by means of the LSZ-reduction formalism. The reverse
problem, namely of how to reconstruct the entire f\/ield content,
or at least part of it, from the scattering theory is in general
still an outstanding challenge. Besides this classif\/ication
issue there remains also the general question if the operators
which are related to the solutions of $(i)$--$(v)$ are genuinely
local, meaning that they (anti)-commute for space-like separations
with themselves. It has been shown that $(i)$--$(v)$ (see e.g.
\cite{Sm,BFK}) imply
\begin{equation}
^{\rm in}\langle\,\phi\,|\,\left[
\mathcal{O}(x),\mathcal{O}(y)\right]
\,|\,\psi\,\rangle^{in}=0\qquad\text{for}\quad(x-y)^{2}<0 \label{3.22}%
\end{equation}
for all matrix elements. Note that commutation rules such as
(\ref{3.22}) hold if the statistics factors in $(i)$--$(v)$ are
trivial, $\sigma^{\mathcal{O}}=1$. However, for more general
statistics factors there are also fermionic, anyonic and even
order-disorder commutation rules as for example for the scaling
$Z_{N}$-Ising model (see \cite{BFK}).

\subsubsection*{Examples of operators and their $\boldsymbol{p}$-functions for the
sinh-Gordon model}

Since there is no backward scattering the Bethe ansatz in
(\ref{3.18}) is trivial, the integrations can be performed
\cite{BK2} and the n-particle $K$-function may be written as
\begin{equation}
K_{n}^{\mathcal{O}}(\underline{\theta})=\sum_{l_{1}=0}^{1}\dots\sum_{l_{n}%
=0}^{1}(-1)^{\sum l_{i}}\prod_{i<j}\!\left(
1+(l_{i}-l_{j})\frac{i\sin\pi\nu }{\sinh\theta_{ij}}\right)
\!p_{n}^{\mathcal{O}}(\underline{\theta
},\underline{l}). \label{3.24}%
\end{equation}

For several cases the correspondence between local operators and
their p-functions have been proposed in
\cite{BK2}\footnote{Sinh-Gordon form factors have been presented
before in a dif\/ferent form in \cite{FMS,KM}.}. Here we provide
three examples:

\begin{enumerate}\itemsep=0pt
\item The {normal ordered exponential of the f\/ield (see also
\cite{BL})}
\begin{equation}
\mathcal{O}(x)=\,:\!e^{\gamma\varphi(x)}\!:\quad\leftrightarrow
p(\underline {\theta},\underline{l})=\left(
\frac{2}{F(i\pi)\sin\pi\nu}\right)  ^{\frac
{n}{2}}\prod\limits_{i=1}^{n}e^{\pi\nu\frac{\gamma}{\beta}(-1)^{l_{i}}},
\label{3.26}%
\end{equation}

\item Expanding the last relation with respect to $\gamma$ one
obtains the $p$-func\-tions for {normal ordered powers
}$:\!\varphi(x)^{N}\!:$ in particular
for $N=1$%
\begin{equation}
\,:\!\varphi(x)\!:\ \leftrightarrow \
p(\underline{\theta},\underline {l})=\frac{\pi\nu}{\beta}\left(
\frac{2}{F(i\pi)\sin\pi\nu}\right)
^{\frac{n}{2}}\sum_{i=1}^{n}(-1)^{l_{i}}, \label{3.28}%
\end{equation}
which yields (for $n=1$) the `wave function renormalization
constant'
\[
Z^{\varphi}=\langle\,0\,|\,\varphi(0)\,|p\,\rangle^{2}=\frac{8\pi^{2}\nu^{2}%
}{-F^{SHG}(i\pi)\beta^{2}\sin\pi\nu}%
\]
(see also \cite{KW}).

\item The higher conserved currents $J_{L}^{\mu}(x)$ which are
typical for integrable quantum f\/ield theories
\[
J_{L}^{\pm}(x)\leftrightarrow\pm
N_{n}^{(J_{L})}\sum_{i=1}^{n}e^{\pm\theta
_{i}}\sum_{i=1}^{n}e^{L\left(  \theta_{i}-\frac{i\pi}{2}(1-(-1)^{l_{i}}%
\nu)\right)  }.
\]
\end{enumerate}

\subsubsection*{Quantum sinh-Gordon f\/ield operator equation}

Using (\ref{3.26}) and (\ref{3.28}) one f\/inds \cite{BK1,BK,BK2}
that the quantum sinh-Gordon f\/ield equation
\begin{equation}
\square\varphi(x)+\frac{\alpha}{\beta}:\!\sinh\beta\varphi\!:(x)=0
\label{3.30}%
\end{equation}
holds for all matrix elements, if the ``bare'' mass
$\sqrt{\alpha}$ is related to the renormalized mass by
\begin{equation}
\alpha=m^{2}\frac{\pi\nu}{\sin\pi\nu} \label{3.32}%
\end{equation}
where $m$ is the physical mass of the fundamental boson. The
result may be checked in perturbation theory by Feynman graph
expansions. In particular in lowest order the relation between the
bare and the renormalized mass (\ref{3.32}) had already been
calculated in the original article \cite{KW}. The result is
\[
m^{2}=\alpha\left(  1-\frac{1}{6}\left(
\frac{\beta^{2}}{8}\right) ^{2}+O(\beta^{6})\right)
\]
which agrees with the exact formula above. The factor
$\frac{\pi\nu}{\sin \pi\nu}$ in (\ref{3.32}) modif\/ies the
classical equation and has to be considered as a quantum
correction. The proof of the quantum f\/ield equation (\ref{3.30})
can be found in \cite{BK2}.

\section{Wightman functions}

As the simplest case we consider the two-point function of two
local scalar
operators $\mathcal{O}(x)$ and $\mathcal{O}^{\prime}(x)$%
\[
w(x)=\langle\,0\,|\,\mathcal{O}(x)\,\mathcal{O}^{\prime}(0)\,|\,0\,\rangle.
\]

\subsubsection*{Summation over all intermediate states}

Inserting a complete set of in-states we may write%
\begin{gather}
w(x)    =\sum_{n=0}^{\infty}\frac{1}{n!}\int\frac{dp_{1}}{2\pi2\omega_{1}%
}\cdots\int\frac{dp_{n}}{2\pi2\omega_{n}}e^{-ix(p_{1}+\cdots+p_{n})}\nonumber\\
\phantom{w(x)    =}{}
\times\langle\,0\,|\,\mathcal{O}(0)\,|p_{1},\ldots,p_{n}\,\rangle
^{\rm in}\,^{\rm in}\langle p_{n},\ldots,p_{1}\,|\,\mathcal{O}^{\prime}%
(0)\,|\,0\,\rangle\nonumber\\
\phantom{w(x) }{} =\sum_{n=0}^{\infty}\frac{1}{n!}\int
d\theta_{1}\cdots\int d\theta
_{n}e^{-ix\sum p_{i}}g_{n}(\underline{\theta}). \label{4.2}%
\end{gather}
We have introduced the functions%
\begin{gather*}
g_{n}(\underline{\theta})    =\frac{1}{\left(  4\pi\right)  ^{n}}%
\langle\,0\,|\,\mathcal{O}(0)\,|p_{1},\ldots,p_{n}\,\rangle^{\rm
in}\,^{\rm in}\langle
p_{n},\ldots,p_{1}\,|\,\mathcal{O}^{\prime}(0)\,|\,0\,\rangle\\
\phantom{g_{n}(\underline{\theta})}{}  =
\frac{1}{\left(  4\pi\right)  ^{n}}F^{\mathcal{O}}(\theta_{1},\ldots
,\theta_{n})F^{\mathcal{O}^{\prime}}(\theta_{n}+i\pi,\ldots,\theta_{1}
+i\pi).
\end{gather*}
where crossing has been used. In particular we consider
exponentials of a scalar bose f\/ield
\[
\mathcal{O}^{(\prime)}(x)=\,:\!e^{i\gamma^{(\prime)}\varphi(x)}\!:
\]
where $:\dots:$ means normal ordering with respect to the physical
vacuum and which amounts to the condition
\[
\,\langle\,0\,|:\!e^{i\gamma^{(\prime)}\varphi(x)}\!:|\,0\,\rangle=1
\]
and therefore $g_{0}=1$ holds.

\paragraph{The Log of the two-point function}

For specif\/ic operators like exponentials of bose f\/ields it
might be convenient (see below) to consider a resummation of the
sum in (\ref{4.2}). For $g_{0}=1$ we may write (see also
\cite{Sm2})
\begin{gather*}
w(x)    =1+\sum_{n=1}^{\infty}\frac{1}{n!}\int
d\theta_{1}\cdots\int
d\theta_{n}e^{-ix\sum p_{i}}g_{n}(\underline{\theta})\\
\phantom{w(x)}{}  =\exp\sum_{n=1}^{\infty}\frac{1}{n!}\int
d\theta_{1}\cdots\int d\theta _{n}e^{-ix\sum
p_{i}}h_{n}(\underline{\theta}).
\end{gather*}
It is well known that the functions $g_{n}$ and $h_{n}$ are
related by the cummulant formula
\[
g_{I}=\sum_{I_{1}\cup\cdots\cup I_{k}=I}h_{I_{1}}\cdots h_{I_{k}},
\]
where we use the short hand notation
$g_{I}=g_{n}(\theta_{1},\dots,\theta _{n})$ with $I=\left\{
1,\dots,n\right\}  $. The relations for the $g$'s and the $h$'s
may be depicted with ~ $g=\begin{picture}(22,12)
\put(10,5){%
\oval(20,10)}\end{picture}$ ~and~ $h=\begin{picture}(22,12)
\put(0,0){%
\framebox(20,10){}}\end{picture}$ ~as
\[
\unitlength.45mm\begin{picture}(240,32)(0,-2) \footnotesize
\put(20,20){\oval (40,20)[]} \put(10,5){\line(0,1){5}}
\put(8,-2){$1$} \put(20,5){\makebox (0,0){$\dots$}}
\put(30,5){\line(0,1){5}} \put(28,-2){$n$} \put(47,15){=}
\put(60,10){\framebox(30,20){}} \put(65,5){\line(0,1){5}}
\put(63,-2){$1$} \put(75,5){\makebox(0,0){$\dots$}}
\put(85,5){\line(0,1){5}} \put(83,-2){$n$}
\put(96,15){$+~\displaystyle \sum_{i=1}^n$}
\put(120,10){\framebox(20,15){}} \put(125,0){\line(0,1){10}}
\put(127.5,4){.\,.} \put(135,0){\line(0,1){10}}
\put(150,10){\framebox (10,10){}} \put(155,5){\line(0,1){5}}
\put(153,-2){$i$} \put(170,14){$+~\cdots$}
\put(200,10){\framebox(10,10){}} \put(205,5){\line(0,1){5}}
\put(203,-2){$1$} \put(220,15){\makebox (0,0){$\dots$}}
\put(230,10){\framebox(10,10){}} \put(235,5){\line(0,1){5}}
\put(233,-2){$n$} \end{picture}
\]
Thus as examples
\begin{gather*}
g_{1}    =h_{1},\\
g_{12}    =h_{12}+h_{1}h_{2},\\
g_{123}    =h_{123}+h_{12}h_{3}+h_{13}h_{2}+h_{23}h_{1}+h_{1}h_{2}h_{3},\\
  \dots\dots\dots\dots\dots\dots\dots
\end{gather*}

Due to Lorentz invariance it is suf\/f\/icient to consider the
value $x=(-i\tau,0)$. Let $\mathcal{O}(x)$ and
$\,\mathcal{O}^{\prime}(x)$ be scalar operators. Then the
functions $h_{n}(\underline{\theta})$ depend only on the rapidity
dif\/ferences. We use the formula for the modif\/ied Bessel
function of the third kind
\[
i\Delta_{+}(x)=\langle\,0\,|\,\varphi(x)\,\varphi(0)\,|\,0\,\rangle=\frac
{1}{4\pi}\int d\theta e^{-\tau
m\cosh\theta}=\frac{1}{2\pi}K_{0}(m\tau)
\]
to perform one integration
\begin{gather*}
\ln w(x)    =\sum_{n=1}^{\infty}\frac{1}{n!}\int
d\theta_{1}\cdots\int
d\theta_{n}e^{-\tau m\sum\cosh\theta_{i}}h_{n}(\underline{\theta})\\
\phantom{\ln w(x)}{} =2\sum_{n=1}^{\infty}\frac{1}{n!}\int
d\theta_{1}\cdots\int d\theta
_{n-1}h_{n}(\theta_{1},\dots,\theta_{n-1},0)K_{0}(m\tau\xi)
\end{gather*}
with
\[
\xi^{2}=\left(  \sum_{i=1}^{n-1}\cosh\theta_{i}+1\right)
^{2}-\left( \sum_{i=1}^{n-1}\sinh\theta_{i}\right)  ^{2}.
\]

\subsubsection*{Short distance behavior $\boldsymbol{x\rightarrow0}$}

In order to perform the conformal limit of massive models one
investigates the short distance behavior (see
e.g.~\cite{Za1,Ca,Sm2}). For small $\tau$ we use
the expansion of the modif\/ied Bessel function of the third kind and obtain%
\begin{gather*}
\ln w(x)    =-2\sum_{n=1}^{\infty}\frac{1}{n!}\int
d\theta_{1}\cdots\int
d\theta_{n-1}h_{n}(\theta_{1},\dots,\theta_{n-1},0)\\
\phantom{\ln w(x)    =}{}\times\left(  \ln
m\tau+\ln\xi+\gamma_{E}-\ln2+O\left(  \tau^{2}\ln \tau\right)
\right)
\end{gather*}
where $\gamma_{E}=0.5772\dots$ is Euler's or Mascheroni's
constant. Therefore the two-point Wightman function has power-like
behavior for short distances
\[
w(x)\approx C\left(  m\tau\right)
^{-4\Delta}\qquad\text{for}\quad \tau \rightarrow0
\]
where the dimension is given by%
\[
\Delta=\frac{1}{2}\sum_{n=1}^{\infty}\frac{1}{n!}\int
d\theta_{1}\cdots\int
d\theta_{n-1}h_{n}(\theta_{1},\dots,\theta_{n-1,}0)
\]
in case the integrals exist. This is true for the exponentials of
bose f\/ields$\,\mathcal{O}=:\!e^{\gamma\varphi(x)}\!:$ due to the
asymptotic
behavior for $\operatorname{Re}\theta_{1}\rightarrow\infty$%
\begin{gather*}
F_{n}^{\mathcal{O}}(\theta_{1},\theta_{2},\dots)    =F_{1}^{\mathcal{O}%
}(\theta_{1})F_{n-1}^{\mathcal{O}}(\theta_{2},\dots)+O(e^{-\theta_{1}}),\\
g_{n}(\theta_{1},\theta_{2},\dots,\theta_{n})
=g_{1}g_{n-1}(\theta _{2},\dots,\theta_{n})+O(e^{-\left\vert
\theta_{1}\right\vert })
\end{gather*}
(see e.g.~\cite{BK2}). Therefore the functions $h_{n}$ satisfy%
\[
h_{n}(\underline{\theta})=O(e^{-\left\vert \theta_{i}\right\vert }%
)\qquad\text{for}\quad
\operatorname{Re}\theta_{i}\rightarrow\pm\infty.
\]
This follows when we distinguish the variable $\theta_{1}$ in the
relation of the $g$'s and the $h$'s above and reorganize the terms
on the right hand side as follows
\[
g_{I}=\sum_{1\in J\subseteq I}h_{J}g_{I\setminus J}.
\]
The constant $C$ is obtained as
\[
C=\exp\left(  -2\sum_{n=1}^{\infty}\frac{1}{n!}\int
d\theta_{1}\cdots\int
d\theta_{n-1}h_{n}(\theta_{1},\dots,\theta_{n-1},0)\left(
\ln\tfrac{1}{2} \xi+\gamma_{E}\right)  \right)
\]
and it should be related to the vacuum expectation value
$G=\langle \,0\,|\mathcal{O}(x)|\,0\,\rangle_{C}$ in the
`conformal normalization'
\[
C=m^{4\Delta}G^{-2}.
\]
Such vacuum expectation value was calculated in \cite{LZ} for the
sine-Gordon model.

\subsubsection*{Example: the sinh-Gordon model}

\begin{figure}[t]
\begin{center}
\setlength{\unitlength}{0.240900pt}
\ifx\plotpoint\undefined\newsavebox {\plotpoint}\fi
\begin{picture}(1200,720)(0,0) \font\gnuplot=cmr10 at 10pt
\gnuplot
\sbox{\plotpoint}{\rule[-0.200pt]{0.400pt}{0.400pt}}%
\put(140.0,82.0){\rule[-0.200pt]{4.818pt}{0.400pt}}
\put(120,82){\makebox(0,0)[r]{0}}
\put(1159.0,82.0){\rule[-0.200pt]{4.818pt}{0.400pt}}
\put(140.0,148.0){\rule[-0.200pt]{4.818pt}{0.400pt}}
\put(120,148){\makebox(0,0)[r]{0.05}}
\put(1159.0,148.0){\rule[-0.200pt]{4.818pt}{0.400pt}}
\put(140.0,215.0){\rule[-0.200pt]{4.818pt}{0.400pt}}
\put(120,215){\makebox(0,0)[r]{0.1}}
\put(1159.0,215.0){\rule[-0.200pt]{4.818pt}{0.400pt}}
\put(140.0,281.0){\rule[-0.200pt]{4.818pt}{0.400pt}}
\put(120,281){\makebox(0,0)[r]{0.15}}
\put(1159.0,281.0){\rule[-0.200pt]{4.818pt}{0.400pt}}
\put(140.0,348.0){\rule[-0.200pt]{4.818pt}{0.400pt}}
\put(120,348){\makebox(0,0)[r]{0.2}}
\put(1159.0,348.0){\rule[-0.200pt]{4.818pt}{0.400pt}}
\put(140.0,414.0){\rule[-0.200pt]{4.818pt}{0.400pt}}
\put(120,414){\makebox(0,0)[r]{0.25}}
\put(1159.0,414.0){\rule[-0.200pt]{4.818pt}{0.400pt}}
\put(140.0,481.0){\rule[-0.200pt]{4.818pt}{0.400pt}}
\put(120,481){\makebox(0,0)[r]{0.3}}
\put(1159.0,481.0){\rule[-0.200pt]{4.818pt}{0.400pt}}
\put(140.0,547.0){\rule[-0.200pt]{4.818pt}{0.400pt}}
\put(120,547){\makebox(0,0)[r]{0.35}}
\put(1159.0,547.0){\rule[-0.200pt]{4.818pt}{0.400pt}}
\put(140.0,614.0){\rule[-0.200pt]{4.818pt}{0.400pt}}
\put(120,614){\makebox(0,0)[r]{0.4}}
\put(1159.0,614.0){\rule[-0.200pt]{4.818pt}{0.400pt}}
\put(140.0,680.0){\rule[-0.200pt]{4.818pt}{0.400pt}}
\put(120,680){\makebox(0,0)[r]{0.45}}
\put(1159.0,680.0){\rule[-0.200pt]{4.818pt}{0.400pt}}
\put(140.0,82.0){\rule[-0.200pt]{0.400pt}{4.818pt}}
\put(140,41){\makebox(0,0){-1}}
\put(140.0,660.0){\rule[-0.200pt]{0.400pt}{4.818pt}}
\put(244.0,82.0){\rule[-0.200pt]{0.400pt}{4.818pt}}
\put(244,41){\makebox(0,0){-0.9}}
\put(244.0,660.0){\rule[-0.200pt]{0.400pt}{4.818pt}}
\put(348.0,82.0){\rule[-0.200pt]{0.400pt}{4.818pt}}
\put(348,41){\makebox(0,0){-0.8}}
\put(348.0,660.0){\rule[-0.200pt]{0.400pt}{4.818pt}}
\put(452.0,82.0){\rule[-0.200pt]{0.400pt}{4.818pt}}
\put(452,41){\makebox(0,0){-0.7}}
\put(452.0,660.0){\rule[-0.200pt]{0.400pt}{4.818pt}}
\put(556.0,82.0){\rule[-0.200pt]{0.400pt}{4.818pt}}
\put(556,41){\makebox(0,0){-0.6}}
\put(556.0,660.0){\rule[-0.200pt]{0.400pt}{4.818pt}}
\put(659.0,82.0){\rule[-0.200pt]{0.400pt}{4.818pt}}
\put(659,41){\makebox(0,0){-0.5}}
\put(659.0,660.0){\rule[-0.200pt]{0.400pt}{4.818pt}}
\put(763.0,82.0){\rule[-0.200pt]{0.400pt}{4.818pt}}
\put(763,41){\makebox(0,0){-0.4}}
\put(763.0,660.0){\rule[-0.200pt]{0.400pt}{4.818pt}}
\put(867.0,82.0){\rule[-0.200pt]{0.400pt}{4.818pt}}
\put(867,41){\makebox(0,0){-0.3}}
\put(867.0,660.0){\rule[-0.200pt]{0.400pt}{4.818pt}}
\put(971.0,82.0){\rule[-0.200pt]{0.400pt}{4.818pt}}
\put(971,41){\makebox(0,0){-0.2}}
\put(971.0,660.0){\rule[-0.200pt]{0.400pt}{4.818pt}}
\put(1075.0,82.0){\rule[-0.200pt]{0.400pt}{4.818pt}}
\put(1075,41){\makebox(0,0){-0.1}}
\put(1075.0,660.0){\rule[-0.200pt]{0.400pt}{4.818pt}}
\put(1179.0,82.0){\rule[-0.200pt]{0.400pt}{4.818pt}}
\put(1179,41){\makebox(0,0){0}}
\put(1179.0,660.0){\rule[-0.200pt]{0.400pt}{4.818pt}}
\put(140.0,82.0){\rule[-0.200pt]{250.295pt}{0.400pt}}
\put(1179.0,82.0){\rule[-0.200pt]{0.400pt}{144.058pt}}
\put(140.0,680.0){\rule[-0.200pt]{250.295pt}{0.400pt}}
\put(140.0,82.0){\rule[-0.200pt]{0.400pt}{144.058pt}}
\sbox{\plotpoint}{\rule[-0.500pt]{1.000pt}{1.000pt}}%
\put(1019,640){\makebox(0,0)[r]{1-particle}}
\multiput(1039,640)(20.756,0.000){5}{\usebox{\plotpoint}}
\put(1139,640){\usebox{\plotpoint}}
\put(141,83){\usebox{\plotpoint}}
\multiput(141,83)(12.743,16.383){2}{\usebox{\plotpoint}}
\multiput(162,110)(12.453,16.604){2}{\usebox{\plotpoint}}
\multiput(183,138)(11.784,17.086){2}{\usebox{\plotpoint}}
\put(214.53,183.47){\usebox{\plotpoint}}
\multiput(224,197)(12.173,16.811){2}{\usebox{\plotpoint}}
\multiput(245,226)(11.902,17.004){2}{\usebox{\plotpoint}}
\multiput(266,256)(11.513,17.270){2}{\usebox{\plotpoint}}
\put(297.63,302.62){\usebox{\plotpoint}}
\multiput(307,316)(12.173,16.811){2}{\usebox{\plotpoint}}
\multiput(328,345)(12.173,16.811){2}{\usebox{\plotpoint}}
\put(358.72,386.50){\usebox{\plotpoint}}
\multiput(370,401)(12.354,16.678){2}{\usebox{\plotpoint}}
\multiput(390,428)(13.041,16.147){2}{\usebox{\plotpoint}}
\put(423.11,467.84){\usebox{\plotpoint}}
\multiput(432,478)(13.995,15.328){2}{\usebox{\plotpoint}}
\put(465.46,513.46){\usebox{\plotpoint}}
\put(480.13,528.13){\usebox{\plotpoint}}
\multiput(494,542)(16.132,13.059){2}{\usebox{\plotpoint}}
\put(527.44,568.48){\usebox{\plotpoint}}
\put(544.50,580.26){\usebox{\plotpoint}}
\put(562.30,590.92){\usebox{\plotpoint}}
\put(580.72,600.42){\usebox{\plotpoint}}
\put(600.18,607.62){\usebox{\plotpoint}}
\put(620.16,613.22){\usebox{\plotpoint}}
\put(640.56,617.03){\usebox{\plotpoint}}
\put(661.29,617.99){\usebox{\plotpoint}}
\put(682.00,616.81){\usebox{\plotpoint}}
\multiput(702,613)(19.690,-6.563){2}{\usebox{\plotpoint}}
\put(741.48,598.96){\usebox{\plotpoint}}
\put(759.48,588.71){\usebox{\plotpoint}}
\put(777.17,577.85){\usebox{\plotpoint}}
\put(794.39,566.29){\usebox{\plotpoint}}
\multiput(806,558)(15.759,-13.508){2}{\usebox{\plotpoint}}
\put(841.73,525.97){\usebox{\plotpoint}}
\put(856.35,511.24){\usebox{\plotpoint}}
\multiput(868,499)(13.995,-15.328){2}{\usebox{\plotpoint}}
\put(898.15,465.11){\usebox{\plotpoint}}
\multiput(910,451)(13.350,-15.893){2}{\usebox{\plotpoint}}
\multiput(931,426)(12.354,-16.678){2}{\usebox{\plotpoint}}
\put(962.46,383.72){\usebox{\plotpoint}}
\multiput(972,371)(12.173,-16.811){2}{\usebox{\plotpoint}}
\multiput(993,342)(12.173,-16.811){2}{\usebox{\plotpoint}}
\put(1023.33,299.67){\usebox{\plotpoint}}
\multiput(1035,283)(11.513,-17.270){2}{\usebox{\plotpoint}}
\multiput(1055,253)(11.902,-17.004){2}{\usebox{\plotpoint}}
\multiput(1076,223)(12.173,-16.811){2}{\usebox{\plotpoint}}
\put(1106.44,180.52){\usebox{\plotpoint}}
\multiput(1118,164)(12.064,-16.889){2}{\usebox{\plotpoint}}
\multiput(1138,136)(12.551,-16.531){3}{\usebox{\plotpoint}}
\put(1179,82){\usebox{\plotpoint}}
\sbox{\plotpoint}{\rule[-0.200pt]{0.400pt}{0.400pt}}%
\put(1019,599){\makebox(0,0)[r]{1+2-particle}}
\put(1039.0,599.0){\rule[-0.200pt]{24.090pt}{0.400pt}}
\put(141,83){\usebox{\plotpoint}}
\multiput(141.58,83.00)(0.496,0.643){39}{\rule{0.119pt}{0.614pt}}
\multiput(140.17,83.00)(21.000,25.725){2}{\rule{0.400pt}{0.307pt}}
\multiput(162.58,110.00)(0.496,0.668){39}{\rule{0.119pt}{0.633pt}}
\multiput(161.17,110.00)(21.000,26.685){2}{\rule{0.400pt}{0.317pt}}
\multiput(183.58,138.00)(0.496,0.727){37}{\rule{0.119pt}{0.680pt}}
\multiput(182.17,138.00)(20.000,27.589){2}{\rule{0.400pt}{0.340pt}}
\multiput(203.58,167.00)(0.496,0.692){39}{\rule{0.119pt}{0.652pt}}
\multiput(202.17,167.00)(21.000,27.646){2}{\rule{0.400pt}{0.326pt}}
\multiput(224.58,196.00)(0.496,0.716){39}{\rule{0.119pt}{0.671pt}}
\multiput(223.17,196.00)(21.000,28.606){2}{\rule{0.400pt}{0.336pt}}
\multiput(245.58,226.00)(0.496,0.692){39}{\rule{0.119pt}{0.652pt}}
\multiput(244.17,226.00)(21.000,27.646){2}{\rule{0.400pt}{0.326pt}}
\multiput(266.58,255.00)(0.496,0.753){37}{\rule{0.119pt}{0.700pt}}
\multiput(265.17,255.00)(20.000,28.547){2}{\rule{0.400pt}{0.350pt}}
\multiput(286.58,285.00)(0.496,0.668){39}{\rule{0.119pt}{0.633pt}}
\multiput(285.17,285.00)(21.000,26.685){2}{\rule{0.400pt}{0.317pt}}
\multiput(307.58,313.00)(0.496,0.692){39}{\rule{0.119pt}{0.652pt}}
\multiput(306.17,313.00)(21.000,27.646){2}{\rule{0.400pt}{0.326pt}}
\multiput(328.58,342.00)(0.496,0.643){39}{\rule{0.119pt}{0.614pt}}
\multiput(327.17,342.00)(21.000,25.725){2}{\rule{0.400pt}{0.307pt}}
\multiput(349.58,369.00)(0.496,0.619){39}{\rule{0.119pt}{0.595pt}}
\multiput(348.17,369.00)(21.000,24.765){2}{\rule{0.400pt}{0.298pt}}
\multiput(370.58,395.00)(0.496,0.625){37}{\rule{0.119pt}{0.600pt}}
\multiput(369.17,395.00)(20.000,23.755){2}{\rule{0.400pt}{0.300pt}}
\multiput(390.58,420.00)(0.496,0.546){39}{\rule{0.119pt}{0.538pt}}
\multiput(389.17,420.00)(21.000,21.883){2}{\rule{0.400pt}{0.269pt}}
\multiput(411.58,443.00)(0.496,0.522){39}{\rule{0.119pt}{0.519pt}}
\multiput(410.17,443.00)(21.000,20.923){2}{\rule{0.400pt}{0.260pt}}
\multiput(432.00,465.58)(0.523,0.496){37}{\rule{0.520pt}{0.119pt}}
\multiput(432.00,464.17)(19.921,20.000){2}{\rule{0.260pt}{0.400pt}}
\multiput(453.00,485.58)(0.551,0.495){35}{\rule{0.542pt}{0.119pt}}
\multiput(453.00,484.17)(19.875,19.000){2}{\rule{0.271pt}{0.400pt}}
\multiput(474.00,504.58)(0.625,0.494){29}{\rule{0.600pt}{0.119pt}}
\multiput(474.00,503.17)(18.755,16.000){2}{\rule{0.300pt}{0.400pt}}
\multiput(494.00,520.58)(0.702,0.494){27}{\rule{0.660pt}{0.119pt}}
\multiput(494.00,519.17)(19.630,15.000){2}{\rule{0.330pt}{0.400pt}}
\multiput(515.00,535.58)(0.814,0.493){23}{\rule{0.746pt}{0.119pt}}
\multiput(515.00,534.17)(19.451,13.000){2}{\rule{0.373pt}{0.400pt}}
\multiput(536.00,548.58)(0.967,0.492){19}{\rule{0.864pt}{0.118pt}}
\multiput(536.00,547.17)(19.207,11.000){2}{\rule{0.432pt}{0.400pt}}
\multiput(557.00,559.59)(1.286,0.488){13}{\rule{1.100pt}{0.117pt}}
\multiput(557.00,558.17)(17.717,8.000){2}{\rule{0.550pt}{0.400pt}}
\multiput(577.00,567.59)(1.560,0.485){11}{\rule{1.300pt}{0.117pt}}
\multiput(577.00,566.17)(18.302,7.000){2}{\rule{0.650pt}{0.400pt}}
\multiput(598.00,574.59)(2.269,0.477){7}{\rule{1.780pt}{0.115pt}}
\multiput(598.00,573.17)(17.306,5.000){2}{\rule{0.890pt}{0.400pt}}
\multiput(619.00,579.61)(4.481,0.447){3}{\rule{2.900pt}{0.108pt}}
\multiput(619.00,578.17)(14.981,3.000){2}{\rule{1.450pt}{0.400pt}}
\put(640,581.67){\rule{5.059pt}{0.400pt}}
\multiput(640.00,581.17)(10.500,1.000){2}{\rule{2.529pt}{0.400pt}}
\put(661,581.67){\rule{4.818pt}{0.400pt}}
\multiput(661.00,582.17)(10.000,-1.000){2}{\rule{2.409pt}{0.400pt}}
\multiput(681.00,580.95)(4.481,-0.447){3}{\rule{2.900pt}{0.108pt}}
\multiput(681.00,581.17)(14.981,-3.000){2}{\rule{1.450pt}{0.400pt}}
\multiput(702.00,577.93)(2.269,-0.477){7}{\rule{1.780pt}{0.115pt}}
\multiput(702.00,578.17)(17.306,-5.000){2}{\rule{0.890pt}{0.400pt}}
\multiput(723.00,572.93)(1.560,-0.485){11}{\rule{1.300pt}{0.117pt}}
\multiput(723.00,573.17)(18.302,-7.000){2}{\rule{0.650pt}{0.400pt}}
\multiput(744.00,565.93)(1.135,-0.489){15}{\rule{0.989pt}{0.118pt}}
\multiput(744.00,566.17)(17.948,-9.000){2}{\rule{0.494pt}{0.400pt}}
\multiput(764.00,556.92)(0.967,-0.492){19}{\rule{0.864pt}{0.118pt}}
\multiput(764.00,557.17)(19.207,-11.000){2}{\rule{0.432pt}{0.400pt}}
\multiput(785.00,545.92)(0.814,-0.493){23}{\rule{0.746pt}{0.119pt}}
\multiput(785.00,546.17)(19.451,-13.000){2}{\rule{0.373pt}{0.400pt}}
\multiput(806.00,532.92)(0.702,-0.494){27}{\rule{0.660pt}{0.119pt}}
\multiput(806.00,533.17)(19.630,-15.000){2}{\rule{0.330pt}{0.400pt}}
\multiput(827.00,517.92)(0.618,-0.495){31}{\rule{0.594pt}{0.119pt}}
\multiput(827.00,518.17)(19.767,-17.000){2}{\rule{0.297pt}{0.400pt}}
\multiput(848.00,500.92)(0.554,-0.495){33}{\rule{0.544pt}{0.119pt}}
\multiput(848.00,501.17)(18.870,-18.000){2}{\rule{0.272pt}{0.400pt}}
\multiput(868.00,482.92)(0.498,-0.496){39}{\rule{0.500pt}{0.119pt}}
\multiput(868.00,483.17)(19.962,-21.000){2}{\rule{0.250pt}{0.400pt}}
\multiput(889.58,460.85)(0.496,-0.522){39}{\rule{0.119pt}{0.519pt}}
\multiput(888.17,461.92)(21.000,-20.923){2}{\rule{0.400pt}{0.260pt}}
\multiput(910.58,438.77)(0.496,-0.546){39}{\rule{0.119pt}{0.538pt}}
\multiput(909.17,439.88)(21.000,-21.883){2}{\rule{0.400pt}{0.269pt}}
\multiput(931.58,415.43)(0.496,-0.651){37}{\rule{0.119pt}{0.620pt}}
\multiput(930.17,416.71)(20.000,-24.713){2}{\rule{0.400pt}{0.310pt}}
\multiput(951.58,389.53)(0.496,-0.619){39}{\rule{0.119pt}{0.595pt}}
\multiput(950.17,390.76)(21.000,-24.765){2}{\rule{0.400pt}{0.298pt}}
\multiput(972.58,363.45)(0.496,-0.643){39}{\rule{0.119pt}{0.614pt}}
\multiput(971.17,364.73)(21.000,-25.725){2}{\rule{0.400pt}{0.307pt}}
\multiput(993.58,336.37)(0.496,-0.668){39}{\rule{0.119pt}{0.633pt}}
\multiput(992.17,337.69)(21.000,-26.685){2}{\rule{0.400pt}{0.317pt}}
\multiput(1014.58,308.29)(0.496,-0.692){39}{\rule{0.119pt}{0.652pt}}
\multiput(1013.17,309.65)(21.000,-27.646){2}{\rule{0.400pt}{0.326pt}}
\multiput(1035.58,279.09)(0.496,-0.753){37}{\rule{0.119pt}{0.700pt}}
\multiput(1034.17,280.55)(20.000,-28.547){2}{\rule{0.400pt}{0.350pt}}
\multiput(1055.58,249.29)(0.496,-0.692){39}{\rule{0.119pt}{0.652pt}}
\multiput(1054.17,250.65)(21.000,-27.646){2}{\rule{0.400pt}{0.326pt}}
\multiput(1076.58,220.21)(0.496,-0.716){39}{\rule{0.119pt}{0.671pt}}
\multiput(1075.17,221.61)(21.000,-28.606){2}{\rule{0.400pt}{0.336pt}}
\multiput(1097.58,190.29)(0.496,-0.692){39}{\rule{0.119pt}{0.652pt}}
\multiput(1096.17,191.65)(21.000,-27.646){2}{\rule{0.400pt}{0.326pt}}
\multiput(1118.58,161.26)(0.496,-0.702){37}{\rule{0.119pt}{0.660pt}}
\multiput(1117.17,162.63)(20.000,-26.630){2}{\rule{0.400pt}{0.330pt}}
\multiput(1138.58,133.40)(0.498,-0.659){79}{\rule{0.120pt}{0.627pt}}
\multiput(1137.17,134.70)(41.000,-52.699){2}{\rule{0.400pt}{0.313pt}}
\put(170,550){$\Delta(\nu)$}
\end{picture}
\end{center}
\vspace{-4mm} \caption{Dimension of an exponential of the f\/ield
for the sinh-Gordon
model: 1- and 1+2-particle inter\-mediate state contributions.}%
\label{f1}%
\end{figure}
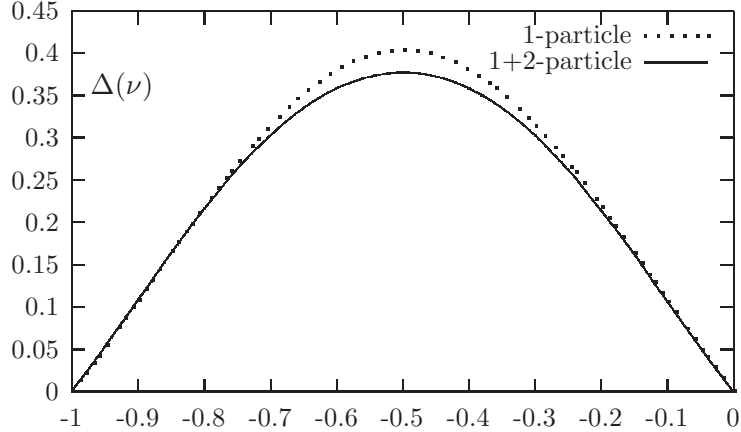

The dimension of the exponential of the f\/ield
$\mathcal{O}(x)=\,:\!e^{\beta \varphi(x)}\!:$ for the sinh-Gordon
model may be calculated in the 1- and 1+2-particle intermediate
state approximation (see Fig.~\ref{f1}) as
\begin{gather*}
\Delta_{1+2}    =\frac{1}{2}\left(  h_{1}+\frac{1}{2!}\int d\theta
\,h_{2}(\theta,0)+\cdots\right) \\
\phantom{\Delta_{1+2}}{}  =-\frac{\sin\pi\nu}{\pi F(i\pi)}+\left(
\frac{\sin\pi\nu}{\pi F(i\pi )}\right)
^{2}\int_{-\infty}^{\infty}d\theta\,\left(  F(\theta)F(-\theta
)-1\right)  +\cdots.
\end{gather*}
The integral may be calculated exactly with the result \cite{BK4}
\[
-\frac{\pi}{2}\sin\pi\nu
F^{2}(i\pi)-\pi\frac{\cos\pi\nu-1}{\sin\pi\nu }+2\left(
1-\frac{\pi\nu\cos\pi\nu}{\sin\pi\nu}\right)  .
\]
In principle the higher particle intermediate state integrals may
also be calculated, however, up to now we were not able to derive
a general formula. For the scaling Ising model, however, this is
possible. The constant $C$ in
the approximation of 1-intermediate states is given as%
\[
C_{1}=\exp\left(  -2h_{1}\left(  \gamma_{E}-\ln2\right)  \right)
=\exp\left( 4\frac{\sin\pi\nu}{\pi F(i\pi)}\left(
\gamma_{E}-\ln2\right)  \right)  .
\]
We have not calculated the integrals appearing in higher-particle
intermediate state contributions.

\subsection*{Acknowledgments}

H.B. was supported partially by the grant  Volkswagenstiftung
within in the project ``Nonperturbative aspects of quantum f\/ield
theory in various space-time dimensions''. H.B. also
acknowled\-ges to ICTP condensed matter group for hospitality,
where part of this work was done. A.F.~acknowledges support from
PRONEX under contract CNPq 66.2002/1998-99 and CNPq (Conselho
Nacional de Desenvolvimento Cient\'{\i}f\/ico e Tecnol\'{o}gico).
This work is also supported by the EU network EUCLID, 'Integrable
models and applications: from strings to condensed matter',
HPRN-CT-2002-00325.

\LastPageEnding

\end{document}